# Self-Similarity of Rogue Wave Generation in Gyrotrons: Beyond the Peregrine Breather

R.M. Rozental *, A.V. Slunyaev, N.S. Ginzburg, A.S. Sergeev, I.V. Zotova

*Institute of Applied Physics of Russian Academy of Sciences,
46 Ul'yanova St., Nizhny Novgorod, 603950, Russia*
* rrz@ipfran.ru

**ABSTRACT**
Within the framework of numerical simulations, we study the gyrotron dynamics under conditions of a significant excess of the operating current over the starting value, when the generation of electromagnetic pulses with anomalously large amplitudes ("rogue waves") can be realized. The averaged shape of high-power pulses is shown to be very close to the celebrated Peregrine breather. At the same time, we demonstrate that the relation between peak power and duration of rogue waves is self-similar, but does not reproduce the one characteristic for Peregrine breathers. Remarkably, the discovered self-similar relation corresponds to the exponential nonlinearity of an equivalent Schrödinger-like evolution equation. This interpretation can be used as a theoretical basis for explaining the giant amplitudes of gyrotron rogue waves.

## I. INTRODUCTION

Gyrotrons are powerful vacuum devices that generate high-frequency radiation based on cyclotron-resonant interaction with a beam of rotating electrons guided by a uniform magnetic field [1]. A feature of gyrotrons, which are a kind of cyclotron resonance masers [2,3], is the operation at the near cutoff frequency of one of the waveguide modes. Alongside with steady-state generation, gyrotrons can exhibit complex dynamic behavior, including chaotic [4,5,6,7]. In particular, as shown theoretically in [8,9], in the regime of developed turbulence, gyrotrons can generate an irregular sequence of ultra-short electromagnetic pulses with a peak power significantly (hundreds of times) higher than the background radiation level. Such random powerful spikes seem to be similar to the so-called rogue waves, i.e., rare events with extremely large amplitudes, and are treated hereafter as such.

Being first discovered in hydrodynamics [10,11], rogue waves have been studied theoretically and registered experimentally in a number of other physical systems [12,13,14]. It may be emphasized that most of the basic mathematical models of rogue waves are based on equations integrable by the inverse scattering technique, which are obtained under the approximations of weak (quadratic or cubic) nonlinearity and weak dispersion. The cubic focusing nonlinear Schrödinger equation (NLSE) seems to be the best known. Rogue waves are most frequently associated with breather-type exact solutions of the NLSE [15]. The similarity of rogue waves occurring in various realms with the Peregrine breather solution [16] has been repeatedly noted by researchers, e.g. [17,18,19,20,21,22,23,24]. The Peregrine breather describes the amplification of wave amplitude as much as thrice; it describes analytically the modulational instability of a uniform wave train with respect to, formally, infinitely long perturbation (in this limit the exponential perturbation growth reduces to a rational dependence). Based on the theoretical analysis of several integrable frameworks, a conclusion was drawn in [25,26], which may be considered as a general conjecture, that anomalously high waves on the background of constant amplitude waves can only form in cases unstable to long modulations (see further discussion in [27]).

Extremely high and short-wave pulses obviously fall beyond the formal limits of applicability of weakly nonlinear frameworks. Hence, the question naturally arises about the rogue wave appearance in strongly nonlinear regimes. Generally speaking, one may anticipate that higher-order nonlinearity can both enhance the rogue wave effect (for example, in very steep sea waves) as well as restrict it (as in breaking ocean waves). Therefore, rogue waves beyond the weakly nonlinear theory may form a less universal picture.

In this paper, we study the characteristics of gyrotron's rogue waves, based on a statistical analysis of the simulated temporal realizations of the output radiation obtained by integrating the self-consistent equations of the electron-wave interaction [8]. As is discovered, the peak power and the duration of the generated extreme spikes are related by a certain relationship, i.e., the process of formation of gyrotron rogue waves has a pronounced self-similar character. Besides, we demonstrate that the discovered self-similar relation corresponds to the exponential nonlinearity of a parabolic Schrödinger-like evolution equation, which can be assigned to the gyrotron output signal, based on the scaling properties of large peaks. This observation is the main finding of the present work. The exponential law of the "equivalent" nonlinear evolution equation provides with a formal explanation of the occurrence of short pulses with peak amplitudes much exceeding the celebrated Peregrine breathers, typical of media with cubic nonlinearity. At the same time, the averaged shape of normalized gyrotron rogue waves agrees surprisingly well with the Peregrine breather solution.

## II. SIMULATIONS OF ROGUE WAVE GENERATION IN GYROTRONS

We simulate the complex dynamics of a gyrotron within the following approximations. The electrodynamic system represents a section of a regular cylindrical waveguide with a cut-off neck on the cathode side (Fig. 1). The gyrotron is driven by a tubular weakly relativistic beam of rotating electrons guided by a uniform magnetic field $\vec{H} = \vec{z}_0 H_0$.

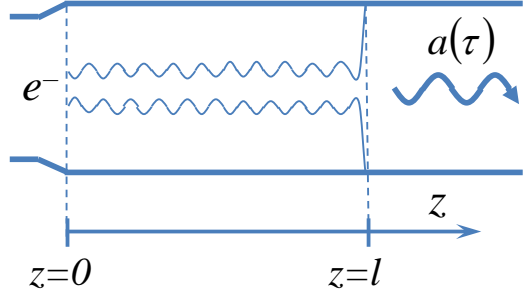

Fig.1. The considered model of the electron-wave interaction in gyrotrons.

The electron-wave interaction takes place at the fundamental-harmonic cyclotron resonance $\omega_c \approx \omega_H$, where $\omega_H = eH_0/mc\gamma$ is the relativistic gyrofrequency, $\gamma$ is the Lorentz factor, $\omega_c$ is the cutoff frequency of a rotating $TE_{mq}$ waveguide mode. The transverse electric field of the excited mode can be presented as [28]

$$\vec{E}_\perp = \kappa^{-1} \mathrm{Re}\left(A(z,t)[\nabla_\perp \Psi \times \vec{z}_0]e^{i\omega_c t}\right), \tag{1a}$$

where $A(z,t)$ is the slowly-varying complex amplitude, $\Psi(r,\varphi) = J_m(\kappa r)e^{-im\varphi}$ describes the field distribution in a circular waveguide, $J_m(x)$ is the Bessel function, $\varphi$ is the azimuthal angle, $\kappa = \omega_c/c$. Then, based on the Ampere-Maxwell equation, the transverse magnetic field of the working mode has the form

$$\vec{H}_\perp = \kappa^{-2} \mathrm{Re}\left(i\frac{\partial A(z,t)}{\partial z}\nabla_\perp \Psi e^{i\omega_c t}\right). \tag{1b}$$

As a result, taking into account the weakly relativistic energy of electrons, the gyrotron dynamics can be described by the following self-consistent system of equations [8], which includes the parabolic equation for the field evolution supplemented by averaged motion equations for electrons:

$$\begin{aligned}
i\frac{\partial^2 a}{\partial Z^2} + \frac{\partial a}{\partial \tau} &= i\frac{I_0}{2\pi}\int_0^{2\pi}\frac{\hat{p}_\perp}{\hat{p}_\|}d\theta_0, \\
\frac{\partial \hat{p}_\perp}{\partial Z} + \frac{g_0^2}{4}\frac{\partial \hat{p}_\perp}{\partial \tau} + i\frac{\hat{p}_\perp}{\hat{p}_\|}\left(\Delta - 1 + |\hat{p}_\perp|^2 + \frac{\hat{p}_\|^2 - 1}{g_0^2}\right) &= i\frac{a}{\hat{p}_\|} + \frac{\beta_{\perp 0}^2}{2}\frac{\partial a}{\partial Z}, \\
\frac{\partial \hat{p}_\|}{\partial Z} + \frac{g_0^2}{4}\frac{\partial \hat{p}_\|}{\partial \tau} &= -g_0^2\frac{\beta_{\perp 0}^2}{2}\mathrm{Re}\left(\frac{\hat{p}_\perp^*}{\hat{p}_\|}\frac{\partial a}{\partial Z}\right),
\end{aligned} \tag{2}$$

In Eqs.(2), the following dimensionless variables and parameters are used:

$$\tau = \omega_c t \frac{\beta_{\perp 0}^4}{8\beta_{\parallel 0}^2}, \quad Z = \frac{\beta_{\perp 0}^2}{2\beta_{\parallel 0}} \frac{\omega_c}{c} z, \quad a = \frac{eAJ_{m-1}(2\pi r_b/\lambda)}{mc\omega_c \gamma_0 \beta_{\perp 0}^3}$$

are the normalized time, coordinate and the RF field amplitude, respectively; $\hat{p}_\perp = e^{-i\omega_c t + i(m-1)\varphi}(p_x + ip_y)/p_{\perp 0}$ and $\hat{p}_\parallel = p_\parallel / mV_{\parallel 0}\gamma_0$ are the normalized transverse and axial electron momenta;

$$I_0 = 16 \frac{eI_b}{mc^3} \frac{\beta_{\parallel 0}}{\gamma_0 \beta_{\perp 0}^6} \frac{J_{m-1}^2(\kappa r_b)}{(\nu_q^2 - m^2) J_m^2(\nu_q)} \tag{3}$$

is the excitation factor; $I_b$ is the current of the tubular electron beam with an injection radius $r_b$; $\nu_q$ is the $q$th root of the Bessel function derivative; $g_0 = \beta_{\perp 0}/\beta_{\parallel 0}$ is the pitch-factor of electrons at the entrance of the interaction space, $\beta_{\perp 0} = V_{\perp 0}/c$ and $\beta_{\parallel 0} = V_{\parallel 0}/c$ are the initial transverse and axial electrons velocities normalized at the speed of light; $\Delta = 2(\omega_c - \omega_H)/\omega_c \beta_{\perp 0}^2$ is the initial cyclotron resonance detuning. In our normalizations, the output power of the gyrotron can be found as

$$P(\tau) = P_{beam} \frac{g_0^2}{1+g_0^2} \frac{2}{I_0} \text{Im}\left( a(Z,\tau) \frac{\partial a(Z,\tau)^*}{\partial Z} \right), \tag{4}$$

where $P_{beam}$ is the power of the electron beam.

When writing the boundary conditions for the motion equations, we assume that, in the cross-section $Z = 0$, the electrons are distributed uniformly over the gyration phases $\theta_0$ and have no initial spread of their velocities:

$$p_+(Z=0) = \exp(i\theta_0), \quad \theta_0 \in [0, 2\pi), \quad \hat{p}_\parallel(Z=0) = 1. \tag{5a}$$

For the field amplitude, we use zero boundary condition at the cut-off neck $Z = 0$:

$$a(Z=0) = 0. \tag{5b}$$

At the collector end $Z = L$ (where $L = \beta_{\perp 0}^2 \omega_c l / 2c\beta_{\parallel 0}$ is the normalized length of the interaction space), the ideal matching with the output waveguide is assumed, which is described by the well-known radiation boundary condition [29]:

$$a(L,\tau) + \frac{1}{\sqrt{\pi i}} \int_0^\tau \frac{1}{\sqrt{\tau-\tau'}} \frac{\partial a(L,\tau')}{\partial Z} d\tau' = 0. \tag{5c}$$

Thus, Eqs.(2) together with the boundary conditions (5a)-(5c) formulate the boundary value problem for description of gyrotron operation regimes. In numerical simulations, the instability develops from a small perturbation of the field amplitudes given by the expression: $a(Z, \tau = 0) = a_0 \sin(\pi Z / L)$.

Based on Eqs.(2)-(5), we simulate various dynamic regime of generation for the normalized cavity length of $L = 15$, the pitch-factor of $g_0 = 1.3$, and the initial transverse velocity of $\beta_{\perp 0} = 0.2$. The chosen parameters are typical for gyrotrons with high output efficiency.

To perform the calculations, a numerical scheme was used, described in detail in [30]. The simulation used an equal grid step in coordinate and time, the value of which was equal to $\Delta Z = \Delta \tau = 0.0845$. Test calculations were performed with a mesh, the step of which was reduced by two and four times. Processing of the data obtained showed the stability of all the results discussed in the work.

It is known (see [8]) that with an increase in the current parameter $I_0$, there is a sequential change in the generation regimes from stationary, through periodic self-modulation, to chaotic ones. In the latter case, the time dependences of the output power represent a random set of radiation spikes. However, near the boundary of the chaotic generation zone ($I_0 = 0.1$), the observed regimes are characterized by a small (6-7) ratio of the spikes peak power to the average power level $\langle P \rangle$ where the angle brackets mean averaging over time. At the same time, at $I_0 \geq 3$, this ratio can reach several hundred with rather frequent occurrences of extremely powerful spikes, which are typical for rogue waves of various nature.

As was shown in [8], the formation of rogue waves includes several stages. At the preliminary stage, electromagnetic radiation is associated with the excitation of a backward wave pulse having a fairly narrow spectrum. Upon reflection from the left side of the system, this radiation is partially absorbed by the electron beam; that leads to a dramatic increase in the transverse energy of electrons. Moving through electrons with a large transverse energy another part of such a pulse is effectively amplified while its duration is shortening. This process is accompanied by the pulse front steepening. The appearance of such pulses leads to a significant widening of the output radiation spectrum, which, in fact, is determined by the value of the detuning parameter $\Delta$.

In turn, in work [31] the dynamics of the system was studied in detail when the detuning parameter $\Delta$ changed. I was shown, that the most optimal region for implementing the chaotic dynamics is the region with negative detuning $\Delta$. In gyrotrons, such regimes are achieved at such values of magnetic fields, when the gyrofrequency exceeds the cutoff frequency of the operating mode. As one can see in Fig.2, for $I_0 = 3$, the gyrotron rogue waves with the highest peak power and the highest frequency of occurrence are realized at $\Delta = -0.7$. This regime was chosen for further statistical analysis carried out in Section III. When this parameter moves to the region of positive values, the efficiency of generating an initiating pulse on the backward wave decreases. In turn, this leads to a significant decrease in the amplitude of the pulse on the following wave. A similar situation occurs when the detuning parameter decreases relative to the optimal value (see Fig. 2).

It should also be noted that in work [8], the mechanism of generation of rogue waves in gyrotrons was confirmed within the framework of modeling based on the universal numerical code KARAT [32,33], which implements the calculation of Maxwell's equations together with the equations of motion of charged macroparticles in a three-dimensional coordinate system. This method takes into account many factors inherent in real physical systems. In particular, factors such as the initial scatter of particles in transverse velocities, the influence of space charge, and the finite conductivity of the walls of the electrodynamic system are taken into account. At the same time, calculations based on a three-dimensional model take significantly more time than calculations using equations (2). In the course of further research, we hope to generate a sufficient set of 3D modeling data and compare with the results obtained in this work.

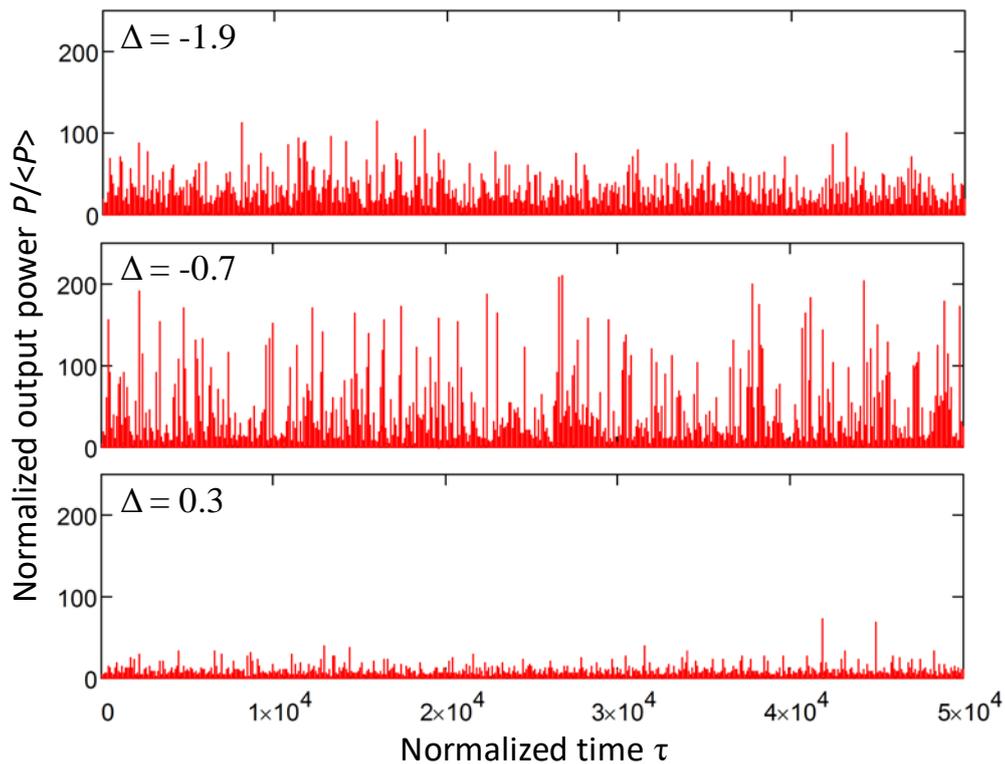

Fig.2. Rogue wave generation in a gyrotron for various value of the cyclotron resonance detuning $\Delta$.

## III. RESULTS OF STATISTICAL ANALYSIS

Note that obtaining time dependences of radiation power like those presented in Fig. 2 requires the integration of the boundary condition (5) from zero normalized time to its current value $\tau$. As a result, for some large $\tau$, the computer resource required for such integration becomes comparable with the resource required to solve the basic Eqs. (2). For optimization of the processes of simulations, we have limited the normalized time for each calculation to $\tau = 50\,000$. In order to obtain a sufficient amount of data, we performed a series of simulations with different values of the initial amplitude $a_0$. A total of ten simulations were performed with the initial amplitude varying from $1\times10^{-6}$ to $10\times10^{-6}$.

For each simulation, we calculated the time dependence of the normalized output power $\rho_i = P_i(\tau)/\langle P_i(\tau)\rangle$, where the brackets $\langle ...\rangle$ designate the averaging over time $\tau$. Then, in the resulting sequence, we select spikes with a peak value of $\rho_{max} > 1$; their width $\Delta\tau$ is determined at the level of $\rho_{max}/2$. Figure 3a shows the "cloud" of ($\Delta\tau$, $\rho_{max}$) pairs found in this way for more than 130 000 spikes in total. One can note the asymmetric horizontal distribution of the density of these pairs. The level curves for the pair's density for the logarithmic-scale horizontal axis are shown in Fig.3b. As expected, most spikes have low power values. In Fig.3c, over the "cloud" of the ($\Delta\tau$, $\rho_{max}$) pairs, we plot the mathematical expectation and the standard deviation of $\Delta\tau$ values, which are calculated using horizontal slices of data on the graph, containing at least 1000 events in each. The standard deviation is calculated separately for the events with duration longer and shorter than its mathematical expectation. It can be seen that the spread of the values changes with increase of $\rho_{max}$; namely, the standard deviation of spikes durations is approximately the same for low power ($\rho_{max} < 10$) and decreases for more rare and extreme events. Since there are a large number of very long spikes with relatively small power, the position of the mathematical expectation of the spike duration is shifted to the right with respect to the most probable values for a given power, which can be seen from a comparison of Figs. 3b and Fig. 3c.

The frequency of extreme groups occurrence is described by the exceedance probability function plotted in Fig. 4, which specifies the probability to meet a spike with the maximum power equal or greater than the given $\rho_{max}$. The distribution is found to be well represented by the power-law dependence close to $\rho_{max}^{-4/3}$ in the interval of relatively small spikes $\rho_{max} < 10$ (the blue dashed line, which is a straight line in the logarithmic scales of Fig. 4a), and by the exponential dependence on the squared power in the interval of very high spikes $\rho_{max} > 70$ (the red dashed-dotted curve, see Fig. 4b). One may see that the distribution at intermediate values of $\rho_{max}$ between the two fits exhibits even slower decay than the power-law approximations provides. The fast decay of the probability distribution function in Fig.4b at very large $\rho_{max}$ may be also a result of the finiteness of the statistical ensemble. Recall that a random superposition of linear waves corresponds to the exponential distribution of the wave power, whereas power-law dependences (so-called heavy tails of the probability distribution) are frequently associated with an abnormally high probability of rare extreme events. Also note that the probability distributions in Fig. 4 do not describe the instantaneous wave power, but concern the peak powers in spikes. The probability distributions for the gyrotron spike power were considered in [8]; which were found to be heavy-tailed.

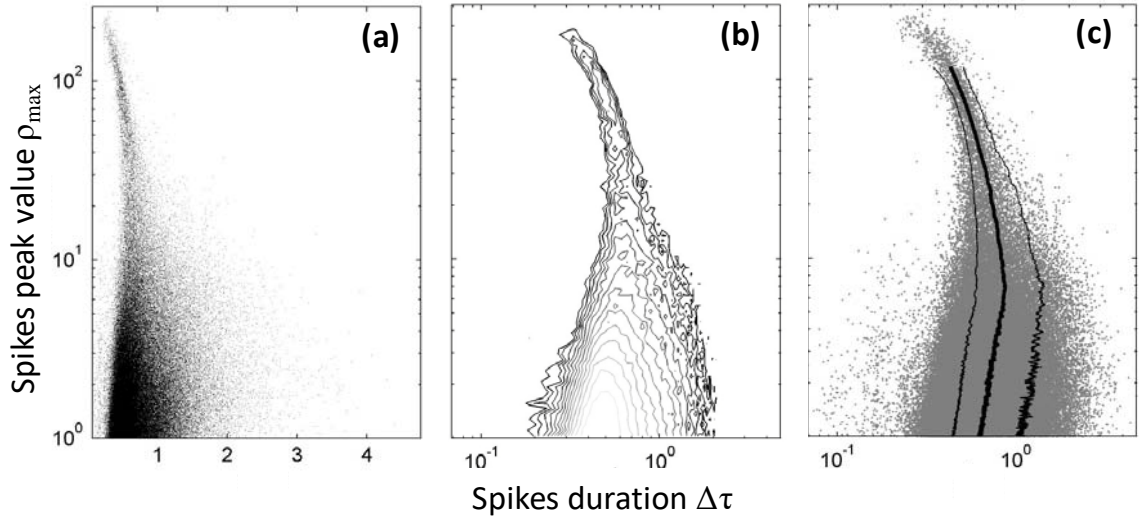

Fig.3. The "cloud" of ($\Delta\tau$, $\rho_{max}$) pairs (a); the density distribution on the parameter plane (b) and the spread of spikes durations for a given peak power (c) (bold curve in the centre is the mathematical expectation, thin lines limit the area within the standard deviations; calculated from samples of 1000 events).

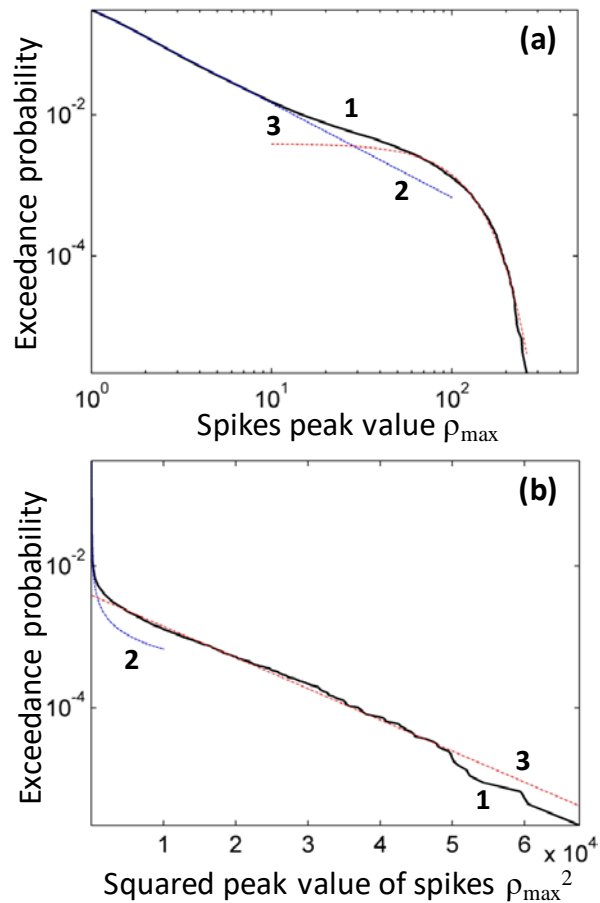

Fig.4. Power exceedance probability distribution (black curve 1) and the power-law ($\sim\rho_{max}^{-1.34}$, the blue dashed curve 2) and exponential ($\sim\exp(-1\cdot10^{-4}\rho_{max}^2)$, the red dashed-dotted curve 3) fits to the data. Panel (a): the data versus $\rho_{max}$ in the logarithmic axes; panel (b): versus $\rho_{max}^2$ in the semilogarithmic axes.

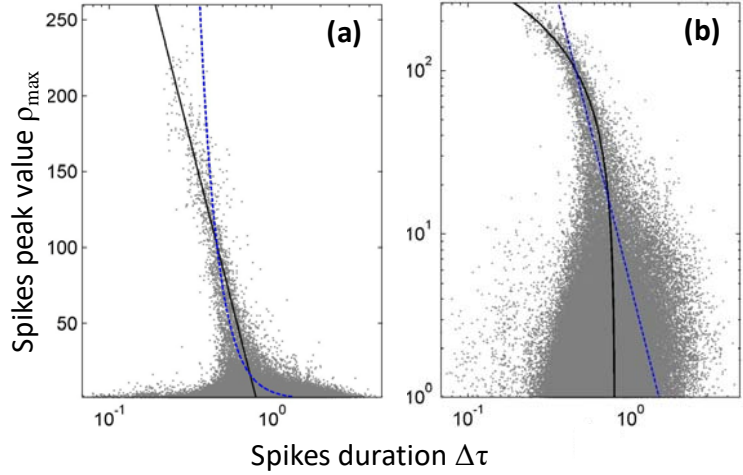

Fig.5. Approximation of data in the region $\rho_{max} > 10$ using logarithmic dependence (solid black line, $a = 181$, $b = 40$) and power law (dashed blue line, $r = 3.9$, $s = 4.9$). Figures (a) and (b) are plotted in semilogarithmic and logarithmic axes, respectively.

According to the data shown in Fig. 3, it becomes obvious that there is a connection between the power and duration of the spikes; which is more distinct for higher-power spikes. From Fig. 3b, it is clear that such a relationship $\rho_{max}(\Delta\tau)$ corresponds to an increase in power with increasing spike duration for at $\rho_{max} \sim< 10$ and with a spike duration decrease for $\rho_{max} >\sim 10$. In the most interesting case of anomalously high-power spikes (rogue waves), the "cloud" of ($\rho_{max}$, $\Delta\tau$) pairs fits well with the logarithmic dependence (solid black lines in Fig.5)

$$\rho_{max} = -a\ln(\Delta\tau) - b, \qquad (6)$$

which in semilogarithmic axes of Fig. 5a represents a straight line. The values of coefficients $a$ and $b$ are chosen so as to minimize the mean-square deviation between the data at $\rho_{max} > 10$ and the fitting curve (6) in the semilogarithmic axes. Note that based on the results of some other numerical experiments, the first parameter $a \approx 200$ shows no significant dependence on the gyrotron parameters $I_0$ and $\Delta$ taken in some relatively broad intervals ($1.5 \leq I_0 \leq 4$ when $\Delta = -0.7$, and $-2 \leq \Delta \leq 0$ when $I_0 = 3$), while the parameter $b$ exhibits strong variation.

The dependence (6) corresponds to a straight line in semilogarithmic axes (solid line in Fig.5a). As an alternative fit for data at $\rho_{max} > 10$, the power-law approximation was considered,

$$\rho_{max} = \frac{s}{(\Delta\tau)^r}, \qquad (7)$$

where the coefficients $r$ and $s$ were evaluated similarly to the previous case, but in logarithmic axes. The corresponding curves are shown by dash in semilogarithmic and logarithmic axes in Fig.5. Any power-law dependence should be straight in Fig.5b; while comparison to the data cloud shows that such an approximation is valid only at small intervals of the spikes power values. The exponential approximation fits the data well at the whole interval of high pulse power.

The characteristic shapes of registered rogue-wave-type spikes are illustrated in Fig. 6. There, the centered shapes $P(\tau)$ of all pulses with $\rho_{max} > 4$ (what is a conventional definition of rogue waves), and also $\rho_{max} > 20$ and $\rho_{max} > 100$, are presented in terms of the their peak powers $P_{max}$ and characteristic durations $\Delta\tau$. The data discrete binning is used in order to estimate the likelihood of wave shapes; the number of cases per bin in logarithmic scale is reflected through the pseudo-color of the contour plots. The averaged curve is determined with maximum accuracy in the vicinity of the maximum, which is used as a reference for the shape centering and normalizing by amplitude. A significant number of large-power pulses are adjacent to even larger bursts; that is why the probability of data exceeding the peak value is not zero at times far from $\tau = 0$. One may conclude that despite significant spread of wave shapes, the normalized data collapses to some universal shape (yellow color). This shape becomes better determined when waves with larger values of $\rho_{max}$ are considered (Fig. 6b,c); for very large $\rho_{max}$ the characteristic shape of gyrotron spikes becomes skewed (Fig. 6c). The blue dashed curve is the analytic solution (10) which we discuss in the following section.

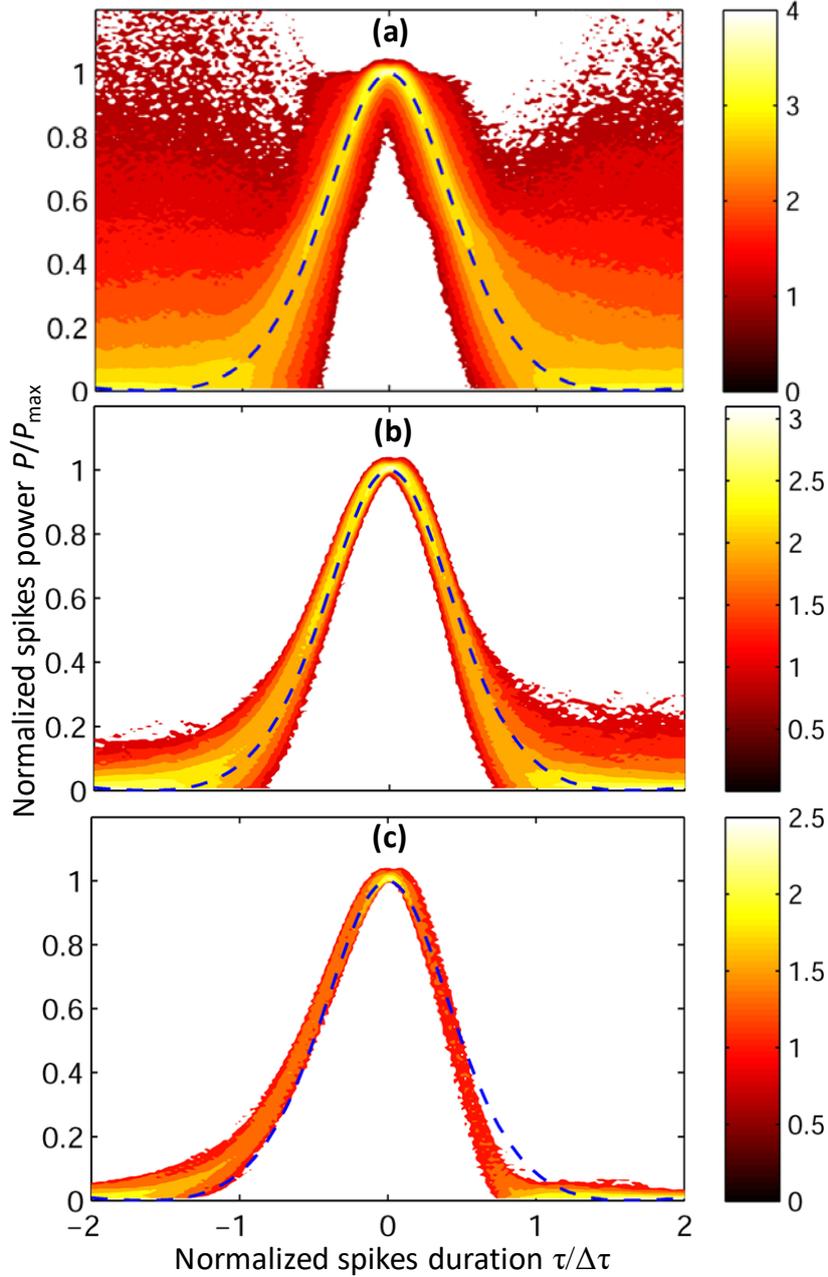

Fig.6. Distribution of rogue wave shapes normalized by their peak powers and characteristic durations for $\rho_{max} > 4$ (a), $\rho_{max} > 20$ (b) and $\rho_{max} > 100$ (c). Colors code the values of the logarithm of the number of spike shapes per bin (up to $\sim 1 \cdot 10^4$ per bin). The blue dashed curve represents the Peregrine solution (10).

## IV. ROGUE WAVE SHAPE SIMILARITY. EVOLUTION EQUATION WITH EQUIVALENT NON-LINEARITY

As discussed in Sec. II, the governing system of equations is complicated, and the internal electrodynamic processes are difficult for the analysis and interpretation. If however, the observed extremely large peaks of output power are associated with a nonlinear instability known as a regular mechanism of rogue wave generation in distributed systems, then it is naturally to assume that the shape of the extreme groups is determined by the process of saturation of the instability growth due to the limiting effect of wave dispersion.

Here we refer to the discussion in the Introduction of the universality of the Peregrine breather shape, observed in numerous examples where the focusing cubic nonlinear Schrödinger equation served as a first approximation for the nonlinear wave evolution. To start, let us consider the example of a cubic self-focusing NLSE on the complex envelope $q(\xi, \tau)$

$$i\frac{\partial q}{\partial \xi} + \alpha |q|^2 q + \beta \frac{\partial^2 q}{\partial \tau^2} = 0. \tag{8}$$

Here $\alpha > 0$ and $\beta > 0$ are real constants and $\xi$ is the evolution variable, while the envelope as a function of time at some coordinate $\xi_0$ is assumed to correspond to the data series under consideration: $\rho(\tau) = |q(\xi_0, \tau)|^2$. The Peregrine breather solution to Eq. (8) may be written in the form [16]

$$q_{PB}(\xi,\tau) = \sqrt{\frac{2}{\alpha}} e^{2i\xi} \left( 1 - 4 \frac{1 + 4i\xi}{1 + 4\frac{\tau^2}{\beta} + 16\xi^2} \right). \tag{9}$$

It tends to a uniform wave solution when $\xi \to \pm\infty$ and represents the maximum wave perturbation with an amplitude three times greater than the background wave when $\xi = 0$. (The solution (9) may be further generalized, but is sufficient for the present purposes). The shape of the Peregrine breather (9) when it attains the maximum amplification may be presented in the normalized scales as follows:

$$\frac{P(\tau)}{P_{\max}} = \left[ \frac{3 - 4\chi^2}{3 + 12\chi^2} \right]^2, \quad \chi = \gamma_P \frac{\tau}{\Delta \tau}, \quad \gamma_P^2 = \frac{3\sqrt{2} - 3}{\sqrt{2} + 3}. \tag{10}$$

Here $P(\tau) = |q_{PB}(0,\tau)|^2$ is the power of the pulse with the maximum value $P_{max} = |q_{PB}(0,0)|^2$, $\Delta\tau$ is the width of the spike at the level of its half power, and the constant $\gamma_P$ is the specific geometric factor of the Peregrine solution.

At the same time, we would like to emphasize that the universality of extreme wave shapes within the NLSE is in fact beyond the particular Peregrine solution, and is a manifestation of the inherent property of the nonlinear system. To illustrate this statement, we plot in Fig. 7 several representative exact solutions $q(\xi, \tau)$ of the NLSE (8), scaled with respect to their maximum amplitudes $A_{\max}$. The plotted solutions are: (i) an envelope soliton (the thick red curve), which is a classic example of a dynamic balance between focusing nonlinearity and spreading dispersion; (ii) few examples of bound states formed by two solitons located at the same point $\tau = 0$ with different combinations of their partial amplitudes (also known as bi-solitons, see [16]), shown at the moment of the maximum positive interference (pale pink curves); and (iii) the Peregrine breather (9) and higher-order rational breathers (so-called super-rogue-waves, see e.g. [20] and references therein) at the instant of the maximum self-modulation (the other curves). Though the small-amplitude "shoulders" of the solutions are clearly different, the solutions represent remarkably similar shapes of extreme wave envelopes (Fig. 7, top). The complex phases of all these solutions are exactly constants (zeros) within the intervals of the main peaks (Fig. 7, bottom), so that all the wave components become in-phase close the envelope maxima.

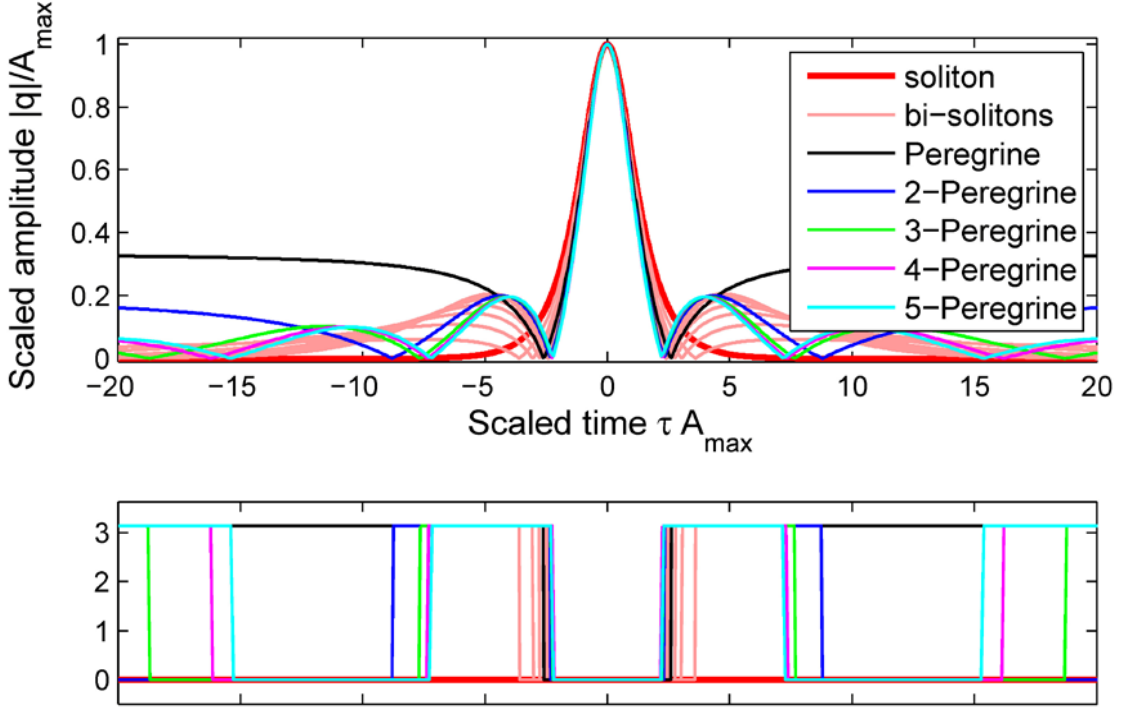

Fig.7. Exact solutions $q(0,\tau)$ of the cubic focusing NLSE in axes scaled using the envelope maximum $A_{max}$: the absolute value (above) and the complex phase in radians (below).

One may note that for the solutions shown in Fig. 7 the dependences of the envelope maxima as functions of the evolution variable (which is the spatial coordinate in our case) are noticeably different. For example, an envelope soliton is characterized by a constant maximum of the envelope, whereas the 5-order Peregrine breather exhibits the most rapid 11-fold wave amplification. Based on this fact, Agafontsev & Gelash have shown that in their numerical simulations of the NLSE most of the largest rogue waves were very well approximated by the amplitude-scaled rational breather solution of the second order (the '2-Peregrine' solution in Fig. 7, which describes 5-fold increase of the wave amplitude) [34]. However, the asymptotic derivation of the NLSE from physical equations implies that the wave envelope evolution is described in the leading order by the advection equation, and hence the duality of the theory for the evolution in space and time takes place, see e.g. examples in [35]. Therefore, in physical applications, the discussed above similarity of the solutions presented in Fig. 7 should be observed to the leading order in both space and time. Thus, it follows that the shape of the Peregrine breather is not specific to rogue waves; in particular, it fits rather well peaks of eternal envelope solitons, when properly scaled. Considering different kinds of rogue-wave-type solutions of the NLSE, Bilman and Miller [36] stated that "all rogue wave solutions having sufficiently large amplitude look the same near the focal point".

The scaling used in Fig. 7, i.e. the relation between the characteristic scales in time and amplitude, $\Delta\tau \sim A_{max}^{-1}$, comes from the similarity parameter of the NLSE (8) which may be introduced proportional to the product of characteristic duration and amplitude of the concerned wave envelope interval. To conclude, the universal shape of solutions in scaled variables, as per Fig. 7, characterizes the governing dynamical system, rather than a specific class of solutions like rogue waves. Remarkably, this shape corresponds well to the averaged shape of gyrotron rogue waves, see the blue dashed curve in Fig. 6 which represents Eq. (10) (note that the power is plotted in Fig. 6 instead of the amplitude in Fig. 7). Thus the normalized shapes of overall gyrotron rogue waves are for some reason very like the shapes of nonlinear structures within the framework of nonlinear-dispersive Schrödinger equation. Some difference between the Peregrine breather shape (10) and the gyrotron waves may be found in Fig. 6b for giant amplifications $\rho_{max} > 20$; if gyrotron rogue waves with even larger amplifications are considered, the difference further grows, but still remains reasonable, see Fig. 6c.

In this section, we consider the gyrotron output signal as a result of evolution of some distributed nonlinear-dispersive system, which dynamics could mimic the characteristics of extreme peaks. Such an equivalent system is constructed in the form of a partial differential equation on wave envelope with two characteristic scales: of the duration $\Delta\tau$ and of the intensity $\rho_{max}$; these scales determine the terms of dispersion and nonlinearity, respectively. We formulate such an abstract evolution equation in the form of a generalized NLSE

$$i\frac{\partial q}{\partial \xi} + \alpha |q|^N q + \beta \frac{\partial^M q}{\partial \tau^M} = 0 \qquad (11)$$

for the complex envelope $q(\xi, \tau)$ of waves propagating in the $M$th order dispersive medium with $(N+1)$th order nonlinearity. For such an equation, the similarity parameter characterizing certain balance between the nonlinear and dispersive terms, yields the condition

$$\left(\rho_{max}\right)^{N/2} \left(\Delta\tau\right)^M = Const. \qquad (12)$$

For instance, for a parabolic equation with cubic nonlinearity $N = 2$ and $M = 2$ (then (11) becomes the cubic nonlinear Schrödinger equation (8)), the similarity parameter is determined by a product of the characteristic amplitude $|q| \sim \rho_{max}^{1/2}$ and the characteristic duration $\Delta\tau$.

In such terms, the relation (6) can be rewritten in the form

$$\exp\left(\rho_{max}\right)\left(\Delta\tau\right)^a = e^{-b}, \qquad (13)$$

which formally corresponds to the exponential nonlinearity of the associated system:

$$i\frac{\partial q}{\partial \xi} + \alpha q \exp\left(\frac{M}{a}|q|^2\right) + \beta \frac{\partial^M q}{\partial \tau^M} = 0. \qquad (14)$$

As well-known, the high order of nonlinearity in the evolution equation can lead to instability of solutions. If the physical energy is determined by the integral

$$\int \rho \cdot d\tau \sim \rho_{max} \Delta\tau \sim \left(\rho_{max}\right)^{1-(N/2)/M}, \qquad (15)$$

(here we use Eq. (12)), then at $N > 2M$, the higher amplitude states become more advantageous in terms of energy. Exponential nonlinearity seemingly might lead to unbounded increase of the wave amplitude and, thus, to numerical instability of simulations. However, such a scenario was not observed in the numerical experiments described in Sect. II: the eventual power amplification is very high, but does not cause instability of the numerical simulations. The stability of the numerical experiments on simulation of such fields must, obviously, arise from internal mechanisms of limiting the generated power within the framework of the integro-differential equations being solved, which have a complicated mathematical structure.

## V. CONCLUSION

Thus, in the operation regimes when rogue waves (i.e., wave pulses with amplitudes much exceeding the average background level) are generated in gyrotrons, these waves demonstrate envelopes which in normalized scales are remarkably close to the famous Peregrine breather. The rogue waves possess self-similarity which can be described by a certain relation between the rogue wave power and its duration. The generated pulses follow the discovered relation in a wide range of the mean power exceedance, from about 20 to 250. Note, that existence of self-similar solutions was also discussed earlier for generation and amplification of short electromagnetic pulses for Cherenkov-type electron-wave interaction in [37,38]. At the same time, the original governing system of equations is so complicated that its reduction to a concise form is so far unobtainable, and thus tangible results are scarce.

Within the equivalent nonlinear-dispersive model, the observed scales of anomalously high-power spikes correspond to the exponential-law nonlinearity, which might be a mathematical justification of giant enhancement of the electromagnetic field power. Examples of evolution equations with complicated (including non-power-law, non-rational) nonlinearity arising in physical problems were discussed earlier in the literature (see, for instance, the references collected in [39,40]). However, we are not aware of the cases when the equation with exponential nonlinearity was encountered.

In this sense, one may say that gyrotron rogue waves possess a unique extremality property. Though the wave amplification described by high-order Peregrine breathers of the NLSE is fundamentally not limited from above (the maximum amplification factor is $2N+1$ for the breather of the $N$-th order [41] where $N$ is a natural number), the probability of occurrence of such waves rapidly decays [42,43], what makes high-order Peregrine breathers practically unrealizable. On the contrary, very high pulses in the simulated gyrotron exhibit high probability of occurrence.

If the wave growth in the registered extreme parcels is balanced by some saturation mechanism (which in this work is associated with the effect of dispersion), the revealed self-similarity relation characterizes the inherent dynamical property of the original physical system. Thus the proposed approach of the statistical analysis of the population of most extreme waves can be used as a diagnostic method for a wide class of non-equilibrium physical systems.

## ACKNOWLEDGMENTS

This work was supported by Science and Education mathematical center of the Lobachevsky Nizhny Novgorod State University under the agreement with Russia's Ministry of Science and Higher Education No. 075-02-2020-1632.

## AUTHOR DECLARATIONS

**Conflict of Interest**

The authors have no conflicts to disclose.

## DATA AVAILABILITY

The data that support the findings of this study are available from the corresponding author upon reasonable request.